\documentclass[%
 amsmath,amssymb,
 aps,
]{revtex4-2}

\usepackage{graphicx}
\usepackage{dcolumn}
\usepackage{bm}

\begin{document}


\title{An acousto-optic modulator based bi-frequency interferometer for quantum technology}

\author{Wenqi Li}
\affiliation{College of Precision Instrument and Opto-Electronics Engineering, Key Laboratory of Opto-Electronics Information Technology, Ministry of Education, Tianjin University, Tianjin 300072, People’s Republic of China\\}
\author{Qiqi Deng}
\affiliation{College of Precision Instrument and Opto-Electronics Engineering, Key Laboratory of Opto-Electronics Information Technology, Ministry of Education, Tianjin University, Tianjin 300072, People’s Republic of China\\}
\author{Xueshi Guo}
\email{xueshiguo@tju.edu.cn}
\affiliation{College of Precision Instrument and Opto-Electronics Engineering, Key Laboratory of Opto-Electronics Information Technology, Ministry of Education, Tianjin University, Tianjin 300072, People’s Republic of China\\}
\author{Xiaoying Li}
\email{xiaoyingli@tju.edu.cn}
\affiliation{College of Precision Instrument and Opto-Electronics Engineering, Key Laboratory of Opto-Electronics Information Technology, Ministry of Education, Tianjin University, Tianjin 300072, People’s Republic of China\\}

\date{\today}

\begin{abstract}
We demonstrate a high performance AOM based bi-frequency interferometer, which can realize either beating or beating free interference for single photon level quantum state. Visibility and optical efficiency of the interferometer are $(99.5\pm0.2)\%$ and $(95\pm1)\%$, respectively. The phase of the interferometer is actively stabilized by using dithering phase locking scheme, where the phase dithering is realized by directly driving the AOMs with specially designed electronic signal. We further demonstrate applications of the interferometer in quantum technology, including bi-frequency coherent combination, frequency tuning and optical switching. These result show the interferometer is a versatile device for multiple quantum technologies.
\end{abstract}

\maketitle

\section{\label{sec:level1}Introduction}
Making use of acousto-optic bragg diffraction effect, Acousto-optic modulators (AOMs) are versatile active optical devices that can efficiently change both the frequency and the propagating direction of an optical field in real time \cite{Wade1990}. 
AOMs have been used in various applications, such as stabilizing the output power, reducing the line-width and shaping the spatial profile of a laser beam etc. \cite{TricotF2018RevSciInst, DongJing2015AO, Song2020OL}. 
AOMs can be exploited to realize active optical tweezers and ultra short pulse picking \cite{Bola2020OL, Oliver2015OE}. In addition to the aforementioned applications in classical technology, AOMs are also indispensable for a lot of quantum technologies. 
Taking the advantage of switching with high isolation, they have been used in photon subtraction based Non-Gaussian state generation \cite{Takase2022OE,PhysRevLett.128.200401}, photon triggered homodyne tomography \cite{Kawasaki2022OE} or controlling of a quantum memory \cite{Appel2008PRL}. 
Taking the advantage of introducing frequency shifting to an optical field, they are used to implement optical heterodyning \cite{Michaud2022IEEEJoQE, Okawa2017OE}, to observe a beating signal from single photons \cite{Mathevet2020AmericanJPhys} or to generate a phase locking reference without displacing the quantum state \cite{Arnbak2019OE}. 
Currently, the maximum diffraction efficiency of commercially available AOMs is around 85$\%$, which reduces as the bandwidth of AOM increases, and harsh requirement on the input beam size and driving power has to been fulfilled in order to achieve this efficiency. 
While the efficiency is enough for most classical applications, it is still not high enough for the quantum applications where quantum states have to transmit through the AOMs and keep the high purity.
This impels us to consider the interferometric enhancement of the diffraction effect when using AOMs in quantum technology.

In most application schemes of AOMs, only one output port (either the direct pass beam or the diffracted beam) is used and the other is abandoned \cite{Ma2020OE, Lima2006APL}. However, AOMs are apparently 4-port devices that can be used for bi-frequency beam splitting. 
Recently, AOM is used as a beam-splitter to implement photon heterodyning \cite{Okawa2017OE}. 
More recently, AOM-based interferometers are used to observe the beating signal of single photons \cite{Mathevet2020AmericanJPhys} or to bidirectional routing laser pulses \cite{Liu:22}, where both beam-splitting and beam-combining functions are carried out by AOMs. 
The visibility of the beating signal in Ref. \cite{Mathevet2020AmericanJPhys} is only about $15\%$, far from the ideal value of 1.
Moreover, both experiments make use of co-propagation optical configuration and create either slow beating of about 1 kHz \cite{Mathevet2020AmericanJPhys} or fast beating signal of about 200 MHz \cite{Liu:22}. On one hand, this co-propagation setup and scanning effect of beating signal will reduce influence the phase drifting and make active phase stabilizing less important. On the other hand, beating free and spatially separated interference can be necessary in many application schemes. 
Here we demonstrate an AOM based bi-frequency interferometer (ABI) working in chopped phase locking \cite{Herzog2006IEEE_JQE} mode, which can realize either beating or beating free interference for single photon level quantum state. By carefully designing the optical system and optimizing the mode matching, the visibility of the beating signal is $(99.5\pm0.2)\%$. We also carefully optimize the optical transmission efficiency of the system. 
Especially, with special designed driving signal for the AOM, the modulation for phase locking \cite{Shuhe2019OC} is carried out with the same AOM for beam splitting. This simplifies the optical setup and further enhances the efficiency of the system.
Moreover, we propose the applications of the interferometer in quantum technology, including bi-frequency coherent combination, frequency tuning and optical switching. 
With the support of experimental testing, we show these applications schemes can in principle achieve near perfect efficiency while only about $50\%$ diffraction efficiency is needed for the AOMs. 
This property will merits many applications for it will greatly reduce the demand for the AOM driving and therefore enhance the bandwidth of the system. These characteristics make the interferometer a powerful tool for many quantum technologies.

\begin{figure}[t]
	\centering
	\includegraphics[width=0.75\textwidth]{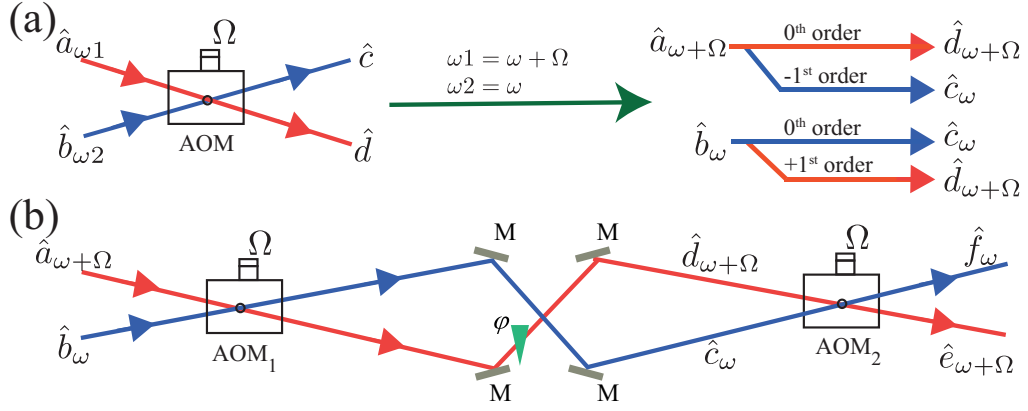}
	\caption{\label{fig1} (a) The optical frequency shifting and interference property when an acousto-optic modulator (AOM) is used as a bi-frequency beam splitter/combiner. (b) An AOM based  bi-frequency interferometer (ABI) whose output ports preserve the optical frequency at the output ports}
\end{figure} 

\section{\label{sec:principle}Working principle of the ABI}
We first illustrate the working principle of the ABI. As shown in Fig. \ref{fig1}(a), without loss of the generality, we assume two optical fields $\hat a_{\omega_1}$ and $\hat b_{\omega_2}$ are aligned to the -1$^{st}$ and the +1$^{st}$ order Bragg diffraction of an AOM, and the output fields $\hat c$ and $\hat d$ have the form of \cite{Mathevet2020AmericanJPhys}
\begin{eqnarray}
	\label{AOM_as_spliiter}
	{\hat c} &=& t  \hat b_{\omega2} + e^{ i\theta  } r  \hat a_{\omega1-\Omega} \nonumber\\
	{\hat d} &=& t  \hat a_{\omega1} - e^{ -i\theta } r  \hat b_{\omega2+\Omega},
\end{eqnarray}
where $\hat a$ and $\hat b$ denote the spatial modes and $\omega_{1(2)}$ in the subscripts denotes their optical frequency. After passing the AOM, the optical frequency is up/down shifted by the frequency of the radio frequency (RF) signal $\Omega$. $t$, $r$ satisfying $t^2+r^2 =1$ are the real number transmission and reflection parameters determined by the strength of the RF signal, and $r^2$ is the diffraction efficiency of the AOM.  $\theta$ is AOM induced phase decided by the phase of the RF signal. Each field operator in Eq. (\ref{AOM_as_spliiter}) satisfies the standard bosonic commutation relation, for example $\hat a$ has the commutation relation $[\hat a, \hat a^\dagger]=1$.

Assuming the frequency of ports $\hat a$ and $\hat b$ are $\omega+\Omega$ and $\omega$, the ouput ports $\hat c$ and $\hat d$ only contains the optical field with frequency of $\omega$ and $\omega+\Omega$, respectively. Therefore, $\hat a_{\omega+\Omega}$ and $\hat b_\omega$ interfere at $\hat c_\omega$ and $\hat d_{\omega+\Omega}$, and can be coherently combined by using another AOM to form a bi-frequency Mach-Zehnder interferometer as shown in Fig. \ref{fig1}(b). 
The output fields of the interferometer can be obtained by using Eq. (\ref{AOM_as_spliiter}) twice:
\begin{eqnarray}
	\label{ABI_as_splitter}
	&{\hat e}_{\omega+\Omega} = t_1' {\hat a}_{\omega+\Omega} - r_1' {\hat b}_{\omega+\Omega} \nonumber\\
	&{\hat f}_{\omega} =  r_2' {\hat a}_{\omega} + t_2' {\hat b}_{\omega}  
\end{eqnarray}
with
\begin{eqnarray}
	\label{ABI_as_splitter2}
	& t'_1 = t_1 t_2 e^{i\varphi} - r_1 r_2 e^{i(\theta_1-\theta_2)} \nonumber\\ 
	& r'_1 = r_1 t_2 e^{i(\varphi-\theta_1)} + t_1 r_2 e^{-i\theta_2} \nonumber\\ 
	& t'_2 = t_1 t_2 - r_1 r_2 e^{i(\theta_2 - \theta_1 +\varphi)}   \nonumber\\
	& r'_2 = r_1 t_2 e^{i\theta_1} + t_1 r_2 e^{i(\theta_2+\varphi)}
\end{eqnarray}
These effective coefficients are decided by the optical path induced phase difference $\varphi$, the AOM splitting parameters $t_{1(2)}$, $r_{1(2)}$, and the AOM driving induced phase $\theta_{1(2)}$ for AOM$_1$ and AOM$_2$, respectively.

Assuming we send light from $\hat b$ port and detect it from $\hat e$ port in Fig. \ref{fig1}(b). Then when the input light is a coherent state $|\alpha\rangle$ and t, r are set to $t_{1(2)}^2=r_{1(2)}^2=0.5$, the intensity at $\hat e$ can be obtained by using Eq. (\ref{ABI_as_splitter})

\begin{equation}
	\label{Eq:I_out}
	\mathrm{I}_{out} = \langle \alpha | {\hat e_{\omega+\Omega}^\dagger}{\hat e_{\omega+\Omega}}| \alpha \rangle = \frac{1}{2}\Big[1+\cos({\phi})\Big] \mathrm{I}_{in},
\end{equation}
where $\phi = \varphi - \theta_1 + \theta_2$ is the overall phase parameter of the interferometer, $\mathrm{I}_{in} = |\alpha|^2$ is the intensity of the input light. 
Eq. (4) describes the ABI in ideal condition. In practice, there inevitably exists the three imperfections: (i) the non-ideal efficiency of the optical components, (ii) the non-ideal mode matching between the two arms of the interferometer , and (iii) the frequency difference between the two RF driving signals. With these factors considered, Eq. (\ref{Eq:I_out}) is rewritten as 
\begin{equation}
	\label{Eq:I_out_eta_V}
	\mathrm{I}_{out}'= \frac{\eta}{2}\Big[1+V\cos({ \Delta \omega T +  \phi})\Big]\mathrm{I}_{in},
\end{equation}
where $\eta$ is the efficiency determined by the losses of the optical elements in ABI, $V$ is the mode mis-matching induced visibility of the interference, and $\Delta \omega T$ is the beating term induced by the frequency difference of the RF signals that drive the two AOMs.  

\section{\label{sec:appli}Applications in quantum technology}

\begin{figure}[h]
	\centering
	\includegraphics[width=0.5\textwidth]{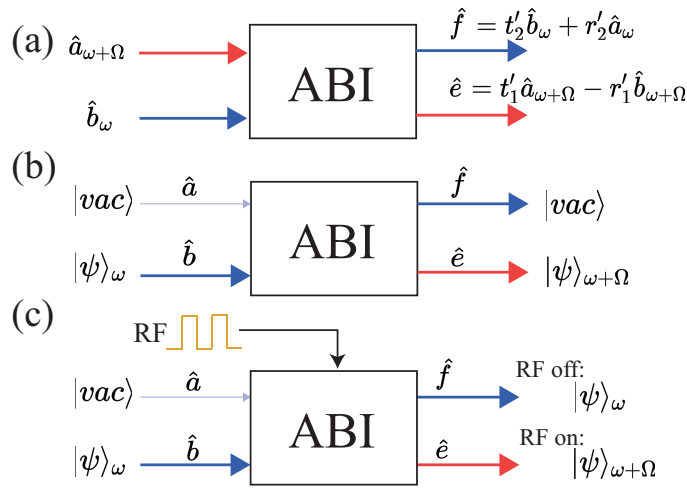}
	\caption{\label{fig_applications} 
		Three typical functions fulfilled by the AOM based bi-frequency interferometer (ABI). (a) A Bi-frequency beam splitter with variable splitting ratio. (b) A frequency tuner of quantum state. (c) An optical switch controlled by gating the RF driving signal to have the AOMs chopped
	}
\end{figure}

To show the ABI with high efficiency and high visibility is a very useful tool for quantum technology, we analyze three typical functions that can be implemented by using ABI (see Fig. \ref{fig_applications}). 
As shown in Fig. \ref{fig_applications}(a), the ABI can be applied to coherently combine two quantum state with a slightly different frequencies. From Eq. (\ref{ABI_as_splitter}-\ref{ABI_as_splitter2}), one sees the ABI can achieve arbitrary effective splitting ratio $t'_{1(2)}$ and $r'_{1(2)}$ when the maximum diffraction efficiency of each AOM can reach $50\%$ and the phase parameters $\varphi$, $\theta_1$ and $\theta_2$ are well controlled. Since frequency is one of the important degree of freedoms for optical mode \cite{Pfister2014PRL}, this scheme can be used to create continuous variable Einstein-Podolsky-Rosen state out of two single mode squeezing state with frequency difference or creating a large scale entangled state with multiple pairs of frequency encoded two mode entangled states \cite{Larsen2019njpQI, Yoshikawa2016APLPhotonics, Asavanant2019science}.

The scheme in Fig. \ref{fig_applications}(a) can interfere two quantum states with different frequency. 
By slightly modifying the scheme (see Fig. \ref{fig_applications}(b)), the ABI can realize a frequency tuner scheme for quantum states, which can tune the frequency of a quantum optical state and fit that of the other quantum system \cite{Suleiman2022QST, Appel2008PRL}. 
In this scheme, a quantum state $|\psi\rangle_\omega$ is sent in at $\hat b$ port and the $\hat a$ port is reserved for vacuum input. According to Eq. (1-4), by controlling the overall phase parameter of the ABI to $\phi = 0$, the quantum state will be shifted by $\Omega$ in optical frequency and output at $\hat e$ port, where $|\psi\rangle_{\omega+\Omega}$ is used to denote the output state with frequency shifted. In this ideal case, the state at $\hat f$ port is a vacuum state. Since this frequency shift is controlled by the frequency $\Omega$ of the RF signal, it provide us a handy tool to realize quantum state frequency tuning. 

If we control the overall phase parameter of the ABI to $\phi = 0$ and $\pi$ in Fig. \ref{fig_applications}(b), the input state $|\psi\rangle_\omega$ will go to port $\hat e$ with frequency shift and port $\hat f$ without frequency shift, respectively. Therefore, the ABI can also be used as an optical switch. However, in this scheme, both the isolation and the transmission efficiency of the switch is confined by the visibility of the interference. In quantum technologies, AOM based switches are often used as a protection stage before a week light detection process \cite{Takase2022OE, Kawasaki2022OE}, and good isolation is a critical requirement for such a switch. To solve this problem, we propose the optical switch scheme in Fig. \ref{fig_applications}(c), where high isolation property is achieved by adding a gate to the RF signals that drive the AOMs. When the RF signals are off, the optical path of the interferometer is degraded into two independent propagated spatial modes. With the same input configuration as Fig. \ref{fig_applications}(b), the input state will go to $\hat f$ port when RF signal is off and the isolation between $\hat e$ and $\hat f$ port is high. When the RF signal is on,  we are able to set the effective transmission/reflection coefficients to $t'_1=0$ and $r'_1=1$ by controlling the overall phase $\phi=0$. In this case, the quantum optical state is transmitted to $\hat e$ port with frequency up shifted.

We note all three schemes in Fig. \ref{fig_applications} can also be implement with a single AOM. However, benefiting from the interferometric enhancement of the diffraction effect, our scheme can achieve a near perfect efficiency when each AOM is working at around 50$\%$ diffraction efficiency. This enhancement is critical in many quantum application schemes, especially for those using continuous variable quantum state. 

\section{\label{sec:Experimental section}Experimental demonstration of the ABI}
\begin{figure}[h]
	\centering
	\includegraphics[width=0.75\textwidth]{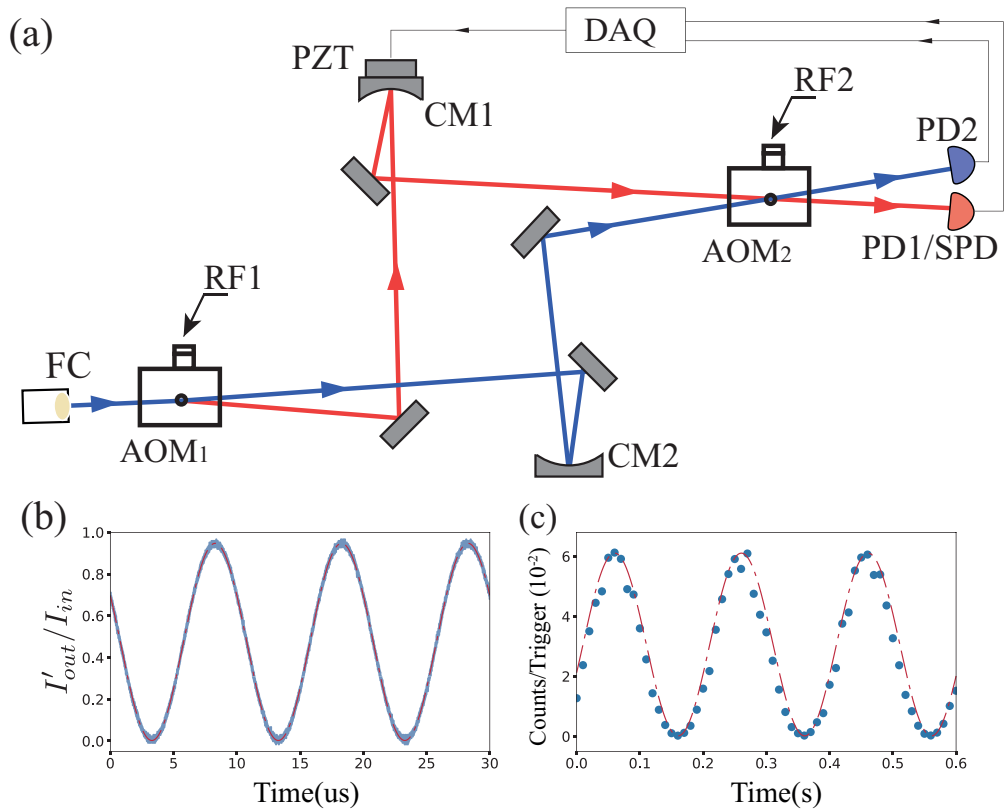}
	\caption{\label{fig_setup} (a) The experimental setup of the AOM based bi-frequency interferometer. 
		AOM, Acousto-optic modulator; CM, concave mirror; PZT, piezo-electric translation device; PD, photo diode; SPD, single photon detector; DAQ, data acquisition system.
		(b) 100 kHz beating signal detected by PD$_1$ from the light with micro-Watt level intensity. (c) 5 Hz Beating signal detected with SPD from the light with single photon level intensity. Red dash-dotted lines in (b) and (c) are the fitting to the data
	}
\end{figure}

Our experimental setup of the acousto-optic modulator based bi-frequency interferometer is shown in Fig. \ref{fig_setup}(a). 
A linearly polarized continuous wave (CW) laser beam with a wavelength of 1550 nm is coupled to the +1$^{st}$ mode ($\hat b_\omega$ in Fig. \ref{fig1}(b)) of the AOMs (ISOMET M1205) with a fiber collimator (FC).
The beam is split into a frequency up-shifted arm (red beam with the frequency of $\omega+\Omega$) and a non-shifted arm (blue beam with the frequency of $\omega$) by AOM$_1$. 
The two beams are then coherently combined by AOM$_2$. For each arm between two AOMs, two plane mirrors and a concave mirror with 300 mm radius of curvature are used to ensure the required mode matching is optimized.  
To control the phase between two arms, a piezo device (PZT) is mounted on the concave mirror of the frequency shifted arm. At the output of the AOM$_2$, 
the frequency up-shifted port is detected by power detector, which is either a photon diode (PD1) for measuring intensity with higher photon number or a single photon detector (SPD, id Quantique-id200) for intensity at single photon level. The output of the frequency non-shifted port is detected with PD$_2$. Detection result for both PD$_1$/SPD and PD$_2$ are sent to a FPGA (STEMlab125-14) based data-acquisition system (DAQ), which also generates the feedback signal to the PZT for phase locking.

\begin{figure}[t]
	\centering
	\includegraphics[width=0.75\textwidth]{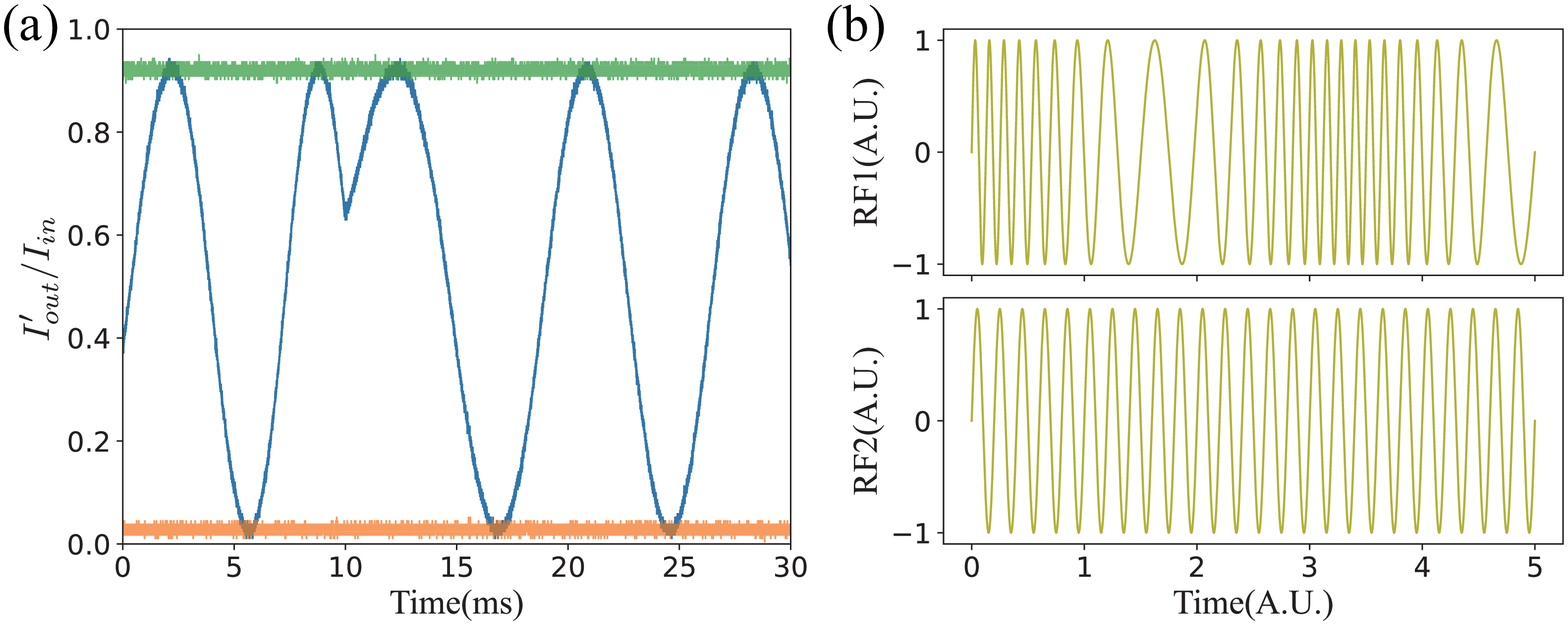}
	\caption{\label{fig_result} 
		(a) Output intensity of an ABI without beating detected by PD$_1$ for PZT scanned (blue), locked to maximum (green) and lock to minimum (orange). 
		(b) The AOM driving signal RF$_1$ and RF$_2$ to realize phase dithering for the phase locking scheme. Both RF$_1$ and RF$_2$ are centered at 80 MHz. The modulation for phase locking of the ABI is realized by adding a 200 kHz small signal phase modulation to RF$_1$. For the illustration propose, the upper trace is plotted with larger modulation strength and frequency than it is used in the experiment
		%
		%
	}
\end{figure}

We first run the setup in slow beating mode, where the active phase stabilizing is off and no signal is added on the PZT. In this mode, we demonstrate the bi-frequency property by the beating effect, and use the beat result to calibrate the visibility $V$ and the transmitting efficiency $\eta$ of the system. RF$_1$ and RF$_2$ drives the AOMs are generated by a two channel arbitrary function generator (Siglent-SDG2122X), and are set to 80 MHz and 79.9 MHz. Thus the beating frequency $\Delta\omega/2\pi$ is 100 kHz. Fig. \ref{fig_setup}(b) shows a measured beating signal for the input light of micro-Watt level intensity, where $\mathrm{I}'_{out}$ is normalized to the intensity of input field $\mathrm{I}_{in}$. The blue trace are the directly measured data and the red dash-dotted line are the fitting to Eq. (5). $V$, $\eta$ and $\phi$ in Eq. (5) are set to be the fitting parameters.
One sees the fitting curve overlaps well with the directly measured data, and the fitting result is V=$(99.5\pm0.2)\%$, $\eta$=$(95  \pm 1)\%$ and $\phi$= (1.08$\pm$0.01) rad, respectively. This result shows we realize almost perfect visibility and high optical efficiency simultaneously. We note each optical surface in our setup has a loss of about $0.5\%$ limited by the quality of optical coating, so $\eta$ can be further improved once components with better optical coatings are used.
We then heavily attenuate the intensity of the input light to single photon level, and detect the beating signal by replacing PD$_1$ with the SPD, which is triggered by its internal 50MHz clock. To get enough counts for a counting window (10 ms), the beating frequency is reduced to 5 Hz by changing the driving frequency of AOM$_2$. The counting result and its fitting are shown in Fig. \ref{fig_setup}(c), where the detection efficiency of the SPD is about $20\%$ and the dark count is $(4\times 10^{-6})$ counts/trigger at this counting setup. Compared with the maximum counting rate of about 0.06 counts/trigger, the dark count is negligible in this measurement. 
Compared to the result in Fig. \ref{fig_setup}(b), which is obtained in the condition of input light at mico-watt level and beating frequency of 100 KHz, data points in Fig. \ref{fig_setup}(c) obviously deviate from the fitting curve due to larger time scale for phase drifting. However, this will not prevent us to get the visibility since it is only determined by the maximum and the minimum value of the curve. The fitting result shows the visibility for light with optical intensity of single photon level is $(99.2\pm0.3)\%$, which is almost the same as the case for input light with higher intensity.


Secondly, we set RF$_1$ and RF$_2$ to be identical frequency of exactly 80 MHz and scan the PZT with a 30 Hz ramp signal.
The interference fringe detected by PD$_1$ is shown by the blue trace in Fig. \ref{fig_result}(a), where the input light is on the  micro-Watt level. When triggered by the ramp signal, one sees the fringe moves on the oscilloscope horizontally in random direction. This is caused by the phase drift between the two arms, and this show it is necessary to introduce active phase locking scheme. We use dithering phase locking scheme \cite{Shuhe2019OC} to stably lock the ABI to arbitrary phase. As it is shown in Fig. \ref{fig_result}(b), modulation required in this phase locking scheme is realized by adding a 200 kHz phase modulation with a small modulation depth to RF$_1$. Therefore, AOM$_1$ is used for both beam splitting and introducing of the modulation for phase locking. This special design avoids using extra optical modulators and enhances the optical efficiency of the system. 
By properly configuring the FPGA system to implement digital demodulators, PID controllers and digital filters \cite{Neuhaus2017CLEO}, the overall phase $\phi$ can be locked to arbitrary value using the detection result of PD$_{1(2)}$.
The green and orange traces in Fig. \ref{fig_result}(a) shows two typical locking results when the overall phase $\phi$ is $0$ and $\pi$ and the values of $\mathrm{I}_{out}/\mathrm{I}_{in}$ at $\hat e$ port is locked to the maximum and minimum point, respectively. 
Compare the beating signal, we find the visibility of interference for the identical frequency case is reduced to $(93.7 \pm 0.5)\%$. Substituting the $\eta$ obtained from the fitting of the beating signal and this visibility into Eq. (\ref{Eq:I_out_eta_V}), we find, the total transmission efficiency of the ABI when used as an optical switch is $\mathrm{I'_{out}/I_{in}=}(92\pm1)\%$. 
We study the reason for this visibility reduction and find, besides a phase shift, the PZT can cause the beam walking-off and induce mode mis-matching. 
This visibility reduction can be in principle avoid if the feedback signal is applied to RF2, and AOM$_2$ is directly used as the phase shifting device of our phase locking system instead of the PZT. 
To complete this optimization, angle calculation and phase unwrapping algorithms is necessary to be implemented in the FPGA \cite{Suleiman2022QST} and the progress is underway.

\begin{figure}[t]
	\centering
	\includegraphics[width=0.75\textwidth]{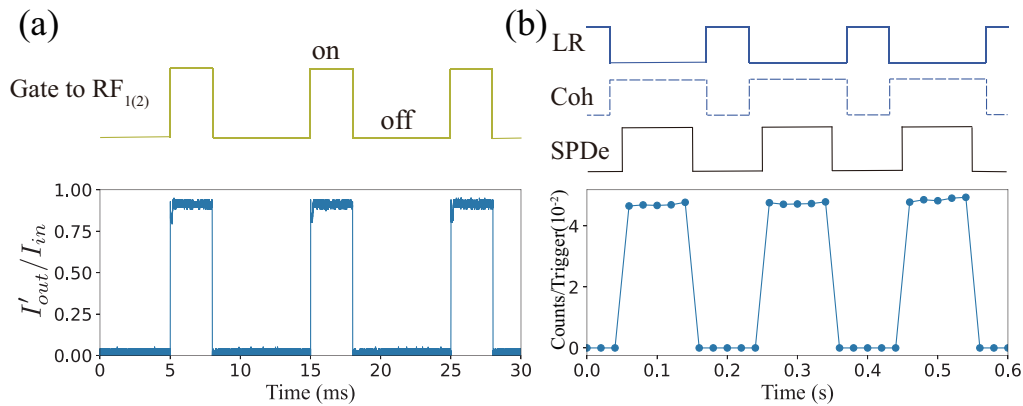}
	\caption{\label{fig_5}  (a) Output intensity in switch mode detected by PD$_1$ with a locking duty cycle of $30\%$. In this mode, the AOM driving signal RF$_{1(2)}$ are turned on and off periodically by a gate signal shown above the intensity result.  
	(b) SPD counting result of the ABI in frequency tuner mode with a detection duty cycle of $50\%$. The error bar for counting result is the same size or smaller than the data point. The input light is a temporally complementary pattern of a locking reference (LR) light with intensity on the micro-Watt level and a week coherent state (Coh) with intensity on the single photon level. The timing sequence of LR and Coh are shown above the counting result. SPDe, SPD trigger enable signal
	}
\end{figure}

In the next, we run the setup in a chopped locking mode for a gated RF signal. In this experiment, the input light is CW with intensity on the micro-Watt level, but the driving signal RF$_1$ and RF$_2$ are gated by two voltage controlled attenuators. This operating mode realizes the optical switch scheme in Fig. \ref{fig_applications}(c).
The RF gate signal is 100 Hz in repetition rate and the duty cycle is tunable.
When the gate signals are on high voltage, the whole system works the same as ABI being locked at the maximum output. When the gate signals are low, the driving signal of the two AOMs in the ABI is completely cut off, and the injected light will go through the frequency non-shifted arm (blue beam) in Fig. \ref{fig_setup}(a), leaving the frequency shifted arm (red beam) in high isolation state to the input field.
A locking feedback enable signal synchronized to the RF gate signal (see the yellow waveform labeled by 'Gate to RF$_{1(2)}$' in Fig. \ref{fig_5}(a)) is send to the DAQ to stop the locking process and hold the feedback value to increase the stability of the locking process.
In the experiment we gradually reduce the duty cycle of the RG signal and find the setup works well with a duty cycle as low as 30$\%$, and a typical measured intensity result detected by PD$_1$ is shown by the blue trace in Fig. \ref{fig_5}(a), in which one sees the lock are stable after a locking establishing time of sub-milli-second order. We note the phase locking stability increases when either the frequency or the duty cycle of the locking process increases.
We further measure the isolation of the ABI when used as an optical switch. Since the isolation exceeds the dynamic range of the SPD, we insert calibrated optical attenuators after the FC when measuring the photon numbers for RF on state and remove it when measuring RF off state. By comparing the photon counting result for the on and off state, we find the isolation is about $74$ dB. This shows the setup functions as a high isolation optical switch from either input port to its double diffracted output port. Though confined by the visibility, isolation of either input port to its double direct pass port is only around 15 dB when used as an optical switch.

Finally, we run the setup in a chopped phase locking mode for a temporally complementary input of a locking reference (LR) light and a week coherent light (Coh). 
The high extinction ratio of $\geq 110$ dB between LR and Coh is ensured by using two mechanical fiber optical switches (e-Photics YFS-2x2), which prevent the SPD from counting the photons in LR.
This mode realizes the frequency tuner scheme in Fig. \ref{fig_applications}(b) for a week input state. The waveform at the top of Fig. \ref{fig_5}(b) shows the timing sequence of this mode, where the micro-Watt level LR is on for $30\%$ of the time and has a frequency of 5 Hz. LR is detected at the frequency non-shifted output port by PD$_2$ and generate feedback signal for phase locking, which is also enabled by a locking feedback enable signal synchronized to LR. When LR is off, the feedback signal is hold and the RF signals continue to drive the AOMs. Coh with a intensity on the single photon level is used as input to simulate a quantum state $|\psi\rangle_\omega$  and detected at the frequency up-shifted port by the SPD. To avoid saturating the SPD, we only enable the SPD using a gate with $50\%$ duty cycle by using a SPD enable signal SPDe. Data points in Fig. \ref{fig_5}(b) shows the counting result with a counting window of 20 ms, where one sees the counting rate stability follows the shape of SPDe signal. This result shows our ABI can work in a chopped locking scheme for single photon level input.

\section{\label{sec:summary}Summary }
In summary, we experimentally realize an AOM based bi-frequency interferometer. The interference visibility of the beating signal of the ABI is (99.5$\pm0.2)\%$ and the optical efficiency is $(95\pm1)\%$.
Benefiting from the interferometric enhancement of the diffraction effect, the interferometer can realize multiple quantum technologies in high efficiency such as bi-frequency coherent combining, optical switching, and frequency tuning.
The interferometer can work in chopped locking mode, which enables the operating of light with intensity on the single photon level. With specially designed RF driving signal, the modulation for phase locking is implemented on the beam splitting AOM, which reduce the number of optical elements and increase the overall quantum efficiency. 
The performance of experimental system we realized is currently confined by the optical loss of the components and the beam walk-off induced by the PZT. With the high visibility of the beating signal, we believe the overall efficiency to implement optical switching or coherently combining of quantum states with the ABI can be further improved to around $98\%$-$99\%$ by using commercially available ultra-low loss optical components.

\begin{acknowledgments}
This work was supported in part by National Natural Science Foundation of China (Grants No.12004279 and 12074283).
\end{acknowledgments}

\bibliography{main}
\end{document}